\documentclass[twocolumn,preprintnumbers,prd,nofootinbib]{revtex4}
\usepackage{graphicx}
\usepackage{times}
\usepackage[hypertex]{hyperref}

\begin{document}

\preprint{UCB-PTH-07/03}
\preprint{LBNL-62350}

\title{Simple Scheme for Gauge Mediation}

\author{Hitoshi Murayama and Yasunori Nomura}
\affiliation{Department of Physics, University of California,
                Berkeley, CA 94720, USA}
\affiliation{Theoretical Physics Group, Lawrence Berkeley National Laboratory,
                Berkeley, CA 94720, USA}

\begin{abstract}
  We present a simple scheme for constructing models that achieve 
  successful gauge mediation of supersymmetry breaking.  In addition 
  to our previous work~\cite{Murayama:2006yf} that proposed drastically 
  simplified models using metastable vacua of supersymmetry breaking in 
  vector-like theories, we show there are many other successful models 
  using various types of supersymmetry breaking mechanisms that rely on 
  enhanced low-energy $U(1)_R$ symmetries.  In models where supersymmetry 
  is broken by elementary singlets, one needs to assume $U(1)_R$ violating 
  effects are accidentally small, while in models where composite fields 
  break supersymmetry, emergence of approximate low-energy $U(1)_R$ 
  symmetries can be understood simply on dimensional grounds.  Even though 
  the scheme still requires somewhat small parameters to sufficiently 
  suppress gravity mediation, we discuss their possible origins due to 
  dimensional transmutation.  The scheme accommodates a wide range of 
  the gravitino mass to avoid cosmological problems.
\end{abstract}
\maketitle

\section{Introduction}
\label{intro}

Despite many new ideas, supersymmetry is still regarded as the prime 
candidate for physics beyond the standard model.  If it exists at 
the TeV scale, it stabilizes the hierarchy between the electroweak 
and the Planck scales, allows for gauge coupling unification with the 
minimal particle content, has a natural candidate for the dark matter, 
and possibly connects to string theory.  On the other hand, having 
been around for three decades, its deficiencies are also well known. 
These include potentially excessive flavor-changing and $CP$-violating 
effects, cosmological gravitino and moduli problems, and the lack of 
automatic proton longevity.  In particular, it has been a nontrivial 
challenge to break supersymmetry and mediate its effect to the 
supersymmetric standard model (SSM) sector in a phenomenologically 
successful manner.

Gauge mediation of supersymmetry breaking~\cite{Dine:1981gu,Dine:1994vc} 
is an attractive solution to the phenomenological problems with 
supersymmetry.  In particular, it naturally avoids excessive 
flavor-changing phenomena because gauge-mediated supersymmetry breaking 
effects are flavor universal.  On the other hand, constructing explicit 
and realistic models of gauge mediation has been a rather nontrivial 
challenge that requires a fair amount of model-building efforts, and 
this aspect has been making the scenario appear a somewhat unlikely 
choice by nature.

In a previous paper~\cite{Murayama:2006yf}, we have proposed a drastically 
simplified class of models for gauge mediation of supersymmetry breaking. 
The models have a supersymmetric $SU(N_c)$, $SO(N_c)$ or $Sp(N_c)$ gauge 
theory with massive quarks, massive vector-like messengers charged under 
the standard model gauge group, and a completely general superpotential 
among these fields.  We have found it remarkable that this simple and 
general class of models can successfully break supersymmetry and generate 
a phenomenologically desired form of supersymmetry breaking masses, 
without any additional ingredients.  This makes us conjecture that gauge 
mediation may be a rather generic phenomenon in the landscape of possible 
supersymmetric theories, which does not require any contrived or artificial 
structures that existed in many of the past models.

In this paper, we show that the success of the previous paper can 
extend more generally to even wider classes of theories.  The low-energy 
structure of the models of Ref.~\cite{Murayama:2006yf} is such that, 
while the entire superpotential does not possess a $U(1)_R$ symmetry, 
terms relevant for supersymmetry breaking possess an accidental 
(and approximate) enhanced $U(1)_R$ symmetry.  In the models of 
Ref.~\cite{Murayama:2006yf}, this structure arises automatically 
at low energies, since $U(1)_R$ violating effects in the supersymmetry 
breaking sector arise from higher dimension operators and thus are 
suppressed by powers of the cutoff scale.  In this paper we present many 
other models that are as simple as those in Ref.~\cite{Murayama:2006yf}, 
and hence the simplicity of the scheme is not necessarily tied to the 
supersymmetry breaking mechanism of Ref.~\cite{Intriligator:2006dd} 
on which the models of Ref.~\cite{Murayama:2006yf} were based.

In addition, in this paper we also consider the possibility that the 
$U(1)_R$ violating terms are suppressed (or absent) without obvious 
low-energy reasons.  Such suppressions may arise through accidentally 
small parameters, as a property of string vacua, or for anthropic reasons. 
We also allow us to make a certain dynamical assumption on the sign 
of a K\"ahler potential term that is not calculable due to strong 
interactions.  These relaxations of the requirements drastically enhance 
a variety of possible theoretical constructions leading to the structure 
described above.  An important key to the success is the mass term for 
the messengers, which is simply one of the generic terms allowed by 
all symmetries.

In order for a model to be viable, several consistency conditions need 
to be met.  Generic gravity-mediated supersymmetry breaking must be 
sufficiently small to avoid excessive flavor-changing and $CP$-violating 
processes.  The impact of $U(1)_R$ violation, both at tree and loop 
levels, must be sufficiently small in the supersymmetry breaking sector 
to keep the essential dynamics intact.  In addition, one should be 
concerned about cosmological constraints on the gravitino, moduli if 
any, the origin of the $\mu$ and $\mu B$ terms, and so on.  Nonetheless, 
the framework we present here is sufficiently general and simple that 
we expect many models can be constructed to address these issues. 
In particular, the framework accommodates a wide range of the gravitino 
mass, $1~{\rm eV} \lesssim m_{3/2} \lesssim 10~{\rm GeV}$.

The simplicity and the variety of the models presented in this paper 
revitalize interest in the gauge mediation scenario, and more generally 
in weak scale supersymmetry.  They largely eliminate the concern about 
weak scale supersymmetry coming from the experimental non-observation 
of flavor-changing or $CP$-violating effects in addition to the ones 
in the standard model.

The organization of the paper is as follows.  In the next section 
we describe the basic framework, and provide general discussions 
that apply to various explicit models presented in later sections. 
In Section~\ref{sec:elementary} we present classes of models in which 
the supersymmetry breaking field is an elementary singlet.  These models 
use accidental features to provide the approximate $U(1)_R$ symmetry. 
In Section~\ref{sec:composite} we present models in which the supersymmetry 
breaking field arises as a composite field.  In these models, $U(1)_R$ in 
the supersymmetry breaking sector arises automatically as an approximate 
low-energy symmetry.  We present classes of models in which the sign 
of the relevant K\"ahler potential term can be reliably calculated 
in the low-energy effective theory, as well as those in which the 
sign is incalculable.  In particular, a class of models presented 
in Section~\ref{subsec:SO10} enjoys the same level of success as the 
one in Ref.~\cite{Murayama:2006yf}.  In Section~\ref{sec:related} 
we discuss slightly different classes of models, which are nonetheless 
closely related to the ones presented in previous sections.  Finally, 
in Section~\ref{sec:retro} we discuss possible ways to naturally 
generate small parameters that are used in the models constructed 
in Sections~\ref{sec:elementary}--\ref{sec:related}.  Conclusions 
are given in Section~\ref{sec:concl}.

\section{Framework}
\label{sec:framework}

In this section we present our basic framework.  Explicit models within 
this framework will be given in later sections.

\subsection{Basic idea}
\label{subsec:basic}

The basic idea is very simple.  We consider the following superpotential
\begin{equation}
  W = - \mu^2 S + \kappa S f \bar{f} + M f \bar{f},
\label{eq:W}
\end{equation}
where $S$ is a gauge singlet chiral superfield (elementary or composite), 
and $f$, $\bar{f}$ are messengers; $\mu$ is the scale of supersymmetry 
breaking, and $\kappa$ a coupling constant.  The parameters $\mu^2$, 
$\kappa$ and $M$ can be taken real and positive without loss of generality. 
For concreteness, we take the messengers to be in ${\bf 5} + {\bf 5^*}$ 
representations of $SU(5)$ in which the standard model gauge group 
is embedded.

We assume that the K\"ahler potential for $S$ takes (approximately) 
the form
\begin{equation}
  K = |S|^2 - \frac{|S|^4}{4 \Lambda^2} 
    + O\left(\frac{|S|^6}{\Lambda^4}\right),
\label{eq:K}
\end{equation}
expanded around the origin $S=0$, where $\Lambda$ is a mass scale.  (We 
assume a canonical K\"ahler potential for the messengers for simplicity.) 
This form of the K\"ahler potential is obtained, for instance, if there 
is an approximate low-energy $U(1)$ symmetry on $S$.  This symmetry can be 
a $U(1)_R$ symmetry possessed by the first two terms of the superpotential, 
Eq.~(\ref{eq:W}), under which $S$, $f$ and $\bar{f}$ carry the charges 
of $2$, $0$ and $0$, respectively.  In this case, the dynamics associated 
with (the generation of) these terms can be responsible for the K\"ahler 
potential of Eq.~(\ref{eq:K}).  (Explicit examples for such dynamics will 
be presented in later sections.)  The $U(1)_R$ symmetry is violated by 
the last term of Eq.~(\ref{eq:W}), but its effect on the K\"ahler potential 
can be suppressed as long as $M \gtrsim \kappa^2 \Lambda/4\pi$, as we 
will see below.

Let us first discuss the model at tree level specified by Eqs.~(\ref{eq:W},%
~\ref{eq:K}).  The potential is simply given by
\begin{eqnarray}
  V &=& \bigl|\mu^2 - \kappa f \bar{f}\bigr|^2 
    \left(1 + \frac{|S|^2}{\Lambda^2} 
    + O\left(\frac{|S|^4}{\Lambda^4}\right) \right)
\nonumber\\
  && {} + \bigl| \kappa S f + M f \bigr|^2 
    + \bigl| \kappa S \bar{f} + M \bar{f} \bigr|^2,
\label{eq:V}
\end{eqnarray}
which has a global supersymmetric minimum at
\begin{equation}
  S = -\frac{M}{\kappa},
\qquad 
  f = \bar{f} = \frac{\mu}{\kappa^{1/2}}.
\label{eq:global-min}
\end{equation}
This potential, however, also has a local supersymmetry breaking minimum 
at the origin of field space, $S = f = \bar{f} = 0$, as long as $M^2 > 
\kappa \mu^2$.  The masses for the scalar components of $S$ and the 
messengers are given by $m_S^2 = \mu^4/\Lambda^2$ and $m_f^2 = M^2 \pm 
\kappa \mu^2$, respectively.  Note that in order for this point to be a 
minimum, it is important that the sign of the second term in Eq.~(\ref{eq:K}) 
is negative.  This can be explicitly proven in some of the models presented 
in later sections, while in some others the sign should be simply assumed.

The tunneling rate from the local minimum to the true supersymmetric 
minimum can be easily suppressed.  To estimate it, we can base our 
discussions on Ref.~\cite{Duncan:1992ai}, calculating a semi-classical 
field theoretic tunneling rate for a toy triangular potential.  While 
the expression worked out there cannot be literally applied to our 
case, we can still approximate our potential by a triangular form, 
obtaining the decay rate per unit volume $\Gamma/V \sim \mu^4 e^{-B}$ 
with $B \sim 8\pi^2 M \Lambda^2/\kappa^{3/2} \mu^3$.  Since $\Lambda$ 
is expected to be $\gtrsim \mu$, the bounce action $B$ can be easily of 
$O(100)$ or larger for $M \gtrsim \kappa^{1/2} \mu$.  To keep the lifetime 
of the local minimum much larger than the age of the universe, we need 
$\Gamma/V \ll H_0^4$, where $H_0 \simeq 1.6 \times 10^{-33}~{\rm eV}$ 
is the present Hubble constant.  The bound is the strongest for $M^2 
\approx \kappa \mu^2$ (the lowest messenger scale), but even then the 
constraints on the parameters are not very strong for $\Lambda \gtrsim M$.

At the local supersymmetry breaking minimum, the messengers $f, \bar{f}$ 
have both supersymmetric and holomorphic supersymmetry breaking masses
\begin{equation}
  M_{\rm mess} = M + \kappa \langle S \rangle 
    \approx M,
\label{eq:M_mess}
\end{equation}
and
\begin{equation}
  F_{\rm mess} = \kappa \langle F_S \rangle 
    = \kappa \mu^2.
\label{eq:F_mess}
\end{equation}
Here, we have assumed that the expectation value of $S$, which is 
generated by $U(1)_R$ violating effects as we will see below, is small. 
The conditions that this requirement imposes on the parameters of 
the theory will be discussed shortly.  The masses for the gauginos 
and scalars in the SSM sector are then generated by messenger 
loops~\cite{Dine:1981gu,Dine:1994vc} and of order
\begin{equation}
  m_{\rm SUSY} \simeq \frac{g^2}{16\pi^2} \frac{\kappa\mu^2}{M},
\label{eq:m_SUSY}
\end{equation}
where $g$ represents generic standard model gauge coupling constants. 
Taking these masses to be of $O(100~{\rm GeV}\!\sim\!1~{\rm TeV})$ 
corresponds to
\begin{equation}
  \frac{\kappa \mu^2}{M} \approx 100~{\rm TeV}.
\label{eq:mess-scale}
\end{equation}
The gravitino mass, on the other hand, is given by
\begin{equation}
  m_{3/2} \approx \frac{\langle F_S \rangle}{M_{\rm Pl}}
    \approx \frac{\mu^2}{M_{\rm Pl}},
\label{eq:m32}
\end{equation}
where $M_{\rm Pl} \simeq 2.4 \times 10^{18}~{\rm GeV}$ is the reduced 
Planck scale.  Thus, requiring that gravity mediation gives only subdominant 
contributions to the scalar masses, $m_{3/2} \lesssim 10~{\rm GeV}$, we find
\begin{equation}
  \mu \lesssim 10^{9.5}~{\rm GeV}.
\label{eq:bound-Lambda}
\end{equation}

\subsection{Effects of {\boldmath $U(1)_R$} violation}
\label{subsec:U1R-vio}

The existence of the supersymmetry breaking minimum at $S = f = \bar{f} = 0$ 
can be viewed as a result of the $U(1)_R$ symmetry possessed by the first 
two terms of Eq.~(\ref{eq:W}): $R(S) = 2$, $R(f) = R(\bar{f}) = 0$.  This 
picture is corrected by $U(1)_R$ violating effects coming from the other 
sectors and/or terms in the theory.  One origin of $U(1)_R$ violation 
arises from the superpotential terms $S^2$ and $S^3$, which are the (only) 
renormalizable terms, other than those in Eq.~(\ref{eq:W}), allowed by 
the gauge symmetry.\footnote{
  A linear term of $S$ in the K\"ahler potential can be absorbed into 
  the definition of the superpotential by the appropriate K\"ahler 
  transformation.}
These terms can be automatically suppressed if $S$ is a composite field 
generated at low energies (as in the models of Section~\ref{sec:composite}) 
but in general must be suppressed for other reasons if $S$ is elementary 
(as in the models of Section~\ref{sec:elementary}).  Denoting the extra 
terms as
\begin{equation}
  \Delta W = \frac{M_S}{2} S^2 + \frac{\kappa_S}{3} S^3,
\label{eq:Delta-W}
\end{equation}
constraints on the parameters $M_S$ and $\kappa_S$ are obtained by 
requiring that the resulting shift of $\langle S \rangle$ is smaller 
than $\approx \Lambda$ (for the expansion of Eq.~(\ref{eq:K}) to be 
valid) and than $\approx M/\kappa$ (to avoid tachyonic messengers):
\begin{equation}
  |M_S| \lesssim {\rm min} \biggl\{ \frac{\mu^2}{\Lambda}, 
    \frac{M \mu^2}{\kappa \Lambda^2} \biggr\},
\qquad
  |\kappa_S| \lesssim \frac{\mu^2}{\Lambda^2}.
\label{eq:bound-MS-kS}
\end{equation}
Note that these conditions are not very restrictive.  This is because 
we use field space with small $S$, where there is a quadratic stabilizing 
potential for $S$ arising from the second term in Eq.~(\ref{eq:K}).

Another source of $U(1)_R$ violation comes from loops of the messengers, 
which do not respect $U(1)_R$ because of the mass term.  These loops 
generate the following Coleman--Weinberg effective potential for $S$:
\begin{eqnarray}
  \Delta V &\approx& \frac{\kappa^2 \mu^4}{16 \pi^2}\, 
    {\cal F}\left(\frac{\kappa S}{M}\right)
\nonumber\\
  &\simeq& \frac{5 \mu^4}{16\pi^2} 
    \left\{ \frac{\kappa^3}{M}(S+S^\dagger) 
    - \frac{\kappa^4}{2 M^2} (S^2+S^{\dagger 2}) 
    + \cdots \right\},
\nonumber\\
\label{eq:Delta-V}
\end{eqnarray}
where ${\cal F}(x)$ is a real polynomial function with the coefficients 
of $O(1)$ up to symmetry factors.  In the second line, we have shown 
the coefficients explicitly, keeping only the leading terms in $\kappa 
\mu^2/M^2$ (and dropping an irrelevant constant in $\Delta V$), which 
corresponds to the correction to the K\"ahler potential of the form 
$\Delta K \approx (1/16\pi^2) \{ (\kappa^3/m)|S|^2(S+S^\dagger) + 
(\kappa^4/m^2)|S|^4 + (\kappa^4/m^2)|S|^2(S^2+S^{\dagger 2}) + \cdots \}$. 
The effective potential of Eq.~(\ref{eq:Delta-V}) pulls the minimum 
at $S=0$ towards the negative direction, and reduces a mass-squared 
eigenvalue of $S$ from $\mu^4/\Lambda^2$.  Yet for
\begin{equation}
  M \gtrsim \frac{\kappa^2}{4\pi} \Lambda,
\label{eq:cond-1}
\end{equation}
we find that these effects are parametrically suppressed and the 
structure of the supersymmetry breaking sector is not significantly 
modified.  In particular, the local minimum stays at small $S$: 
$|\langle S \rangle| \approx \kappa^3 \Lambda^2/16\pi^2 M \lesssim 
{\rm min}\{ M/\kappa, \Lambda \}$.  The condition for avoiding tachyonic 
messengers is
\begin{equation}
  M^2 \gtrsim \kappa \mu^2.
\label{eq:cond-2}
\end{equation}
Note that the inequalities of Eqs.~(\ref{eq:cond-1},~\ref{eq:cond-2}) 
should be understood that order one coefficients are omitted.

In general, the 4 parameters of the theory $\mu$, $\kappa$, $M$ and 
$\Lambda$ are arbitrary, except that we expect $\Lambda \gtrsim \mu$ 
if the higher dimension term in the K\"ahler potential of Eq.~(\ref{eq:K}) 
is induced by the dynamics generating the first (two) term(s) of the 
superpotential of Eq.~(\ref{eq:W}).  By varying these parameters, a wide 
variety of physical pictures can arise.  For $\mu^2/\Lambda \gg M$, for 
example, we can first integrate out the $S$ scalar, which is much heavier 
than the messengers, and then the low-energy theory below the $S$ mass 
appears as the standard gauge mediation model, with the Lagrangian given 
by $\int\!d^2\theta\, (M_{\rm mess} + \theta^2 F_{\rm mess}) f \bar{f} 
+ {\rm h.c.}$  On the other hand, in the opposite limit of $M \gg 
\mu^2/\Lambda$, we can first integrate out the messengers $f$ and 
$\bar{f}$.  This generates ``gaugino mass operators'' $\int\!d^2\theta\, 
S {\cal W}^\alpha {\cal W}_\alpha + {\rm h.c.}$ as well as flavor universal 
``scalar mass operators'' $\int\!d^4\theta\, S^\dagger S \Phi^\dagger \Phi$, 
where ${\cal W}_\alpha$ represents the SSM gauge field strength superfields 
and $\Phi$ the SSM matter and Higgs chiral superfields.  The low-energy 
theory below the messenger mass, $M$, is then a simple Polonyi-type
model -- Eqs.~(\ref{eq:W},~\ref{eq:K}) with $f$ and $\bar{f}$ set to 
zero -- together with these operators, which are responsible for the 
masses of the gauginos and scalars in the SSM sector.

\subsection{The origin of {\boldmath $S$} and the scales of the theories}
\label{subsec:S}

The framework described here represents a great simplification in building 
models of gauge mediation.  The only required aspect of model building is 
essentially to explain the origin of the quartic term in Eq.~(\ref{eq:K}). 
There are many classes of explicit models that can be constructed in this 
framework, some of which will be presented in Sections~\ref{sec:elementary} 
and \ref{sec:composite}. In the models where $S$ is an elementary singlet 
(the models in Section~\ref{sec:elementary}), it must be assumed that the 
$U(1)_R$ violating terms of Eq.~(\ref{eq:Delta-W}) are suppressed without 
obvious low-energy reasons.  On the other hand, in the models where $S$ 
is a composite field (the models in Section~\ref{sec:composite}), these 
terms are naturally suppressed.  Suppose that the $S$ field consists of 
$n$ elementary fields, $S \sim Q^n/\Lambda_s^{n-1}$ ($n \geq 2$), where 
$Q$ and $\Lambda_s$ represent generic constituents of $S$ and the scale 
of compositeness, respectively.  The parameters $M_S$ and $\kappa_S$ in 
Eq.~(\ref{eq:Delta-W}) are then suppressed as
\begin{equation}
  M_S \approx \frac{\Lambda_s^{2n-2}}{M_*^{2n-3}},
\qquad
  \kappa_S \approx \frac{\Lambda_s^{3n-3}}{M_*^{3n-3}},
\label{eq:MS-kS}
\end{equation}
respectively.  Here, $M_*$ is the cutoff scale of the theory.

The $S$ compositeness also suppresses the parameter $\kappa$ in 
Eq.~(\ref{eq:W}), weakening the transmission of gauge mediation effects. 
Writing the fundamental superpotential, which replaces the first two 
terms of Eq.~(\ref{eq:W}), schematically as
\begin{equation}
  W \approx -\frac{\zeta}{M_*^{n-3}}Q^n + \frac{\eta}{M_*^{n-1}} Q^n f \bar{f},
\label{eq:W-schem}
\end{equation}
we find
\begin{equation}
  \mu^2 = \frac{\zeta \Lambda_s^{n-1}}{M_*^{n-3}},
\qquad
  \kappa = \frac{\eta \Lambda_s^{n-1}}{M_*^{n-1}},
\label{eq:corresp}
\end{equation}
where we have defined the compositeness scale $\Lambda_s$ by $S = 
Q^n/\Lambda_s^{n-1}$.  The requirement of Eq.~(\ref{eq:bound-MS-kS}) 
for preserving the approximate $U(1)_R$ symmetry was to have a metastable 
minimum around the origin to justify the analysis.  It requires an 
unexplained suppression in $M_S$ for elementary $S$, while it is easy 
to satisfy for composite $S$.

The gauge-mediated contribution to the SSM superparticles is given 
by (see Eq.~(\ref{eq:m_SUSY}))
\begin{equation}
  m_{\rm SUSY} \simeq \frac{g^2}{16\pi^2} 
    \frac{\zeta \eta \Lambda_s^{2n-2}}{M_*^{2n-4}M_{\rm mess}},
\label{eq:m_SUSY-2}
\end{equation}
where we have denoted the messenger mass explicitly as $M_{\rm mess}$, 
leaving the possibility that the term with the $S$ expectation value 
contributes significantly in Eq.~(\ref{eq:M_mess}).  We find that 
$m_{\rm SUSY}$ is suppressed by $(\Lambda_s/M_*)^{2n-2}$.  The 
contribution from gravity mediation is (see Eq.~(\ref{eq:m32}))
\begin{equation}
  m_{3/2} \approx \frac{\zeta \Lambda_s^{n-1}}{M_*^{n-3} M_{\rm Pl}}.
\label{eq:m32-2}
\end{equation}
Dividing Eq.~(\ref{eq:m32-2}) by Eq.~(\ref{eq:m_SUSY-2}), and using 
the stability condition $M_{\rm mess}^2 \gtrsim \kappa \mu^2$ (see 
Eq.~(\ref{eq:cond-2})), we obtain
\begin{equation}
  \frac{m_{\rm 3/2}}{m_{\rm SUSY}} \approx \frac{16\pi^2}{g^2} 
    \frac{M_*^{n-1} M_{\rm mess}}{\eta \Lambda_s^{n-1} M_{\rm Pl}} 
  \gtrsim 100 \sqrt{\frac{\zeta}{\eta}} \frac{M_*}{M_{\rm Pl}}.
\label{eq:ratio}
\end{equation}
In order for the gauge-mediated contribution to dominate over the 
gravity-mediated one, we must have $m_{3/2}/m_{\rm SUSY} \lesssim 
O(0.01\!\sim\!0.1)$.  This requires either small $\zeta$, large 
$\eta$, small $M_*$, or a combination of these:
\begin{equation}
  \sqrt{\frac{\zeta}{\eta}} \frac{M_*}{M_{\rm Pl}} 
  \lesssim O(10^{-4}\!\sim\!10^{-3}).
\label{eq:gen-bound}
\end{equation}
(A large value for $\eta$ is obtained by generating the nonrenormalizable 
coupling $W \sim Q^n f \bar{f}$ by integrating out heavy fields below 
$M_*$ in the theory.)  In the case that $S$ is a two-body composite, 
i.e. $n=2$, this condition is satisfied simply by having small mass 
parameters for elementary fields: $W \sim m QQ$ with $m \ll M_*$, which 
corresponds to having small $\zeta$.

A large variety of theoretical constructions allowed in this framework 
can lead to a wide range of the parameters $\mu$, $\kappa$, $M$ and $\Lambda$. 
This implies in particular that the framework accommodates a wide range of 
the gravitino mass, $1~{\rm eV} \lesssim m_{3/2} \lesssim 10~{\rm GeV}$. 
The smallest gravitino mass is obtained when $M^2 \approx \kappa \mu^2 
\approx (100~{\rm TeV})^2$ and $\kappa \approx O(1\!\sim\!4\pi)$.  Such 
a light gravitino is useful to avoid cosmological problems associated 
with the gravitino~\cite{Pagels:1981ke}.

\section{Theories with Elementary Singlets}
\label{sec:elementary}

In this section we present classes of models in which the supersymmetry 
breaking field $S$ is an elementary singlet.  As discussed in the previous 
section, this case requires accidental suppressions in $U(1)_R$ violating 
terms.  Nonetheless it is quite nontrivial that successful gauge mediation 
is obtained in very simple models once such suppressions are assumed.

\subsection{Tree-level supersymmetry breaking}
\label{subsec:OR}

An obvious candidate for producing the required K\"ahler 
potential of Eq.~(\ref{eq:K}) is the good-old O'Raifeartaigh 
model~\cite{O'Raifeartaigh:1975pr}.  We replace the first term 
of Eq.~(\ref{eq:W}) (and the second term of Eq.~(\ref{eq:K})) by
\begin{equation}
  W = -\mu^2 S + \lambda S X^2 + m X Y,
\label{eq:W_OR}
\end{equation}
where $S$, $X$ and $Y$ are singlet fields having the canonical K\"ahler 
potential (up to terms suppressed by the cutoff scale).  Here, we 
simply assume that possible terms $S^2$, $S^3$, $X^2$, $SXY$, $Y^2$ 
and $SY^2$ are somehow suppressed.  (The other terms can be forbidden 
by a discrete $Z_2$ symmetry under which $X$ and $Y$ are odd.)  The 
parameters $\mu^2$, $\lambda$ and $m$ are taken real and positive 
without loss of generality.

The superpotential of Eq.~(\ref{eq:W_OR}) breaks supersymmetry due 
to the incompatibility between $F_S = 0$ and $F_Y = 0$.  For $m^2 > 
2 \lambda \mu^2$, the minimum is at $X = Y = 0$.  The field $S$ is a flat 
direction at tree level, but is stabilized at the origin due to radiative 
corrections to the K\"ahler potential.  These corrections can be calculated 
most easily by computing the Coleman-Weinberg effective potential for 
$S$, arising from loops of $X$ and $Y$.  The mass matrix of the $X$ and 
$Y$ fermions in the basis $(\psi_X, \psi_Y)$ is
\begin{equation}
  \left( \begin{array}{cc}
    2\lambda S & m \\
    m          & 0
  \end{array} \right),
\end{equation}
while that of the scalars in the basis $(X, X^\dagger, Y, Y^\dagger)$ is
\begin{equation}
  \left( \begin{array}{cccc}
    m^2 + 4\lambda^2 |S|^2  & -2\lambda\mu^2  &  2\lambda m S^\dagger  &  0 \\
    -2\lambda\mu^2  & m^2 + 4\lambda^2 |S|^2  &  0  &  2\lambda m S         \\
    2\lambda m S     &     0   &  m^2  &  0 \\
    0  &  2\lambda m S^\dagger &  0  &  m^2
  \end{array} \right).
\end{equation}
The resulting Coleman-Weinberg potential can be expanded around the origin 
of $S$ as
\begin{equation}
  \Delta V = \frac{\lambda^4 \mu^4}{3\pi^2 m^2} |S|^2 
    - \frac{3 \lambda^6 \mu^4}{10\pi^2 m^4} |S|^4 + \cdots,
\label{eq:pot_OR}
\end{equation}
where we have dropped an unimportant constant and kept only the leading 
terms in $\lambda \mu^2 / m^2$.  Note that since the superpotential of 
Eq.~(\ref{eq:W_OR}) possesses a $U(1)_R$ symmetry under which $S$, $X$ 
and $Y$ carry the charges of $2$, $0$ and $2$, respectively, the potential 
of Eq.~(\ref{eq:pot_OR}) is a function only of $|S|^2$.  This, therefore, 
corresponds to the K\"ahler potential corrections of the form of 
Eq.~(\ref{eq:K}), with $\Lambda^2 = 3\pi^2 m^2/\lambda^4$.

To summarize, the complete superpotential of the model presented here is 
given by the combination of Eqs.~(\ref{eq:W}) and (\ref{eq:W_OR}):
\begin{equation}
  W = -\mu^2 S + \lambda S X^2 + m X Y 
    + \kappa S f \bar{f} + M f \bar{f}.
\label{eq:complete-OR}
\end{equation}
The other possible renormalizable terms must be suppressed as discussed 
in Section~\ref{subsec:U1R-vio}.  The K\"ahler potential can be canonical.

\subsection{Dynamical models}
\label{subsec:IYIT}

Another class of models that reproduces the super- and K\"ahler potentials 
of Eqs.~(\ref{eq:W},~\ref{eq:K}) uses supersymmetry breaking theories 
of Ref.~\cite{Izawa:1996pk}, based on quantum modified moduli space. 
Consider an $SU(2)$ gauge theory with four doublets $Q_i$ and six singlets 
$S^{ij}=-S^{ji}$ $(i,j=1,\cdots,4)$.  It is convenient to exploit the 
local equivalence of $SU(4)$ and $SO(6)$ groups for the flavor symmetry, 
and regard both singlets $S^{ij}$ and mesons $M_{ij} \equiv Q_i Q_j$ to be 
in the vector representation of $SO(6)$.  For the sake of presentation, we 
assume that flavor $SO(6)$ is explicitly broken to $SO(5)$ by superpotential 
interactions, and refer to $SO(5)$ vectors $S_a$, $M_a$ $(a=1,\cdots,5)$ 
and singlets $S_6$, $M_6$.\footnote{
  The model works equally well if $SO(6)$ is completely broken by 
  superpotential interactions analogous to Eq.~(\ref{eq:W_IYIT}).}
The superpotential, which replaces the first term of Eq.~(\ref{eq:W}) 
and the second term of Eq.~(\ref{eq:K}), is then given by
\begin{equation}
  W = -\lambda_5 S_a M_a - \lambda S_6 M_6.
\label{eq:W_IYIT}
\end{equation}
The couplings $\lambda_5$ and $\lambda$ can be taken real and positive 
without loss of generality.  At quantum level, the theory confines with 
the following quantum modified moduli space~\cite{Seiberg:1994bz}:
\begin{equation}
  {\rm Pf} (Q_i Q_j) = M_a M_a + M_6 M_6 = \Lambda_s^4,
\label{eq:QM-moduli}
\end{equation}
where $\Lambda_s$ is the dynamical scale of $SU(2)$ gauge interactions. 
Because this constraint contradicts with the conditions for a supersymmetric 
vacuum $\partial W/\partial S_a = \partial W/\partial S = 0$, the theory 
breaks supersymmetry.  Assuming $\lambda_5 > \lambda$, the minimum is 
at $M_a = S_a = 0$ and $M_6 = \Lambda_s^2$.  We can thus eliminate $M_6$ 
using the constraint as $M_6 = (\Lambda_s^4 - M_a M_a)^{1/2}$, and the 
superpotential of Eq.~(\ref{eq:W_IYIT}) becomes
\begin{equation}
  W = -\lambda_5 S_a M_a - \lambda S (\Lambda_s^4 - M_a M_a)^{1/2},
\label{eq:W_IYIT_2}
\end{equation}
where we have denoted $S_6$ simply as $S$.  The field $S$ is a flat 
direction at tree level.  We thus need to consider quantum effects 
to find where the minimum is for $S$.  For $S \gg \Lambda_s$, the 
potential grows logarithmically with $S$~\cite{Arkani-Hamed:1997ut}. 
This can be shown explicitly because in this regime a weakly coupled 
description in terms of the fundamental quarks $Q_i$ is valid, so that 
the wavefunction renormalization factor $Z_S$ can be reliably calculated. 
The potential is $V_{\rm eff} = Z_S^{-1}(S)|F_S|^2$, which grows for 
large $S$ because of the Yukawa coupling $\lambda$.

The behavior of the potential for small $S$ is more subtle.  It was 
shown, however, in Ref.~\cite{Chacko:1998si} that the behavior of the 
K\"ahler potential around the origin of $S$ is indeed of the type in 
Eq.~(\ref{eq:K}).  The quartic correction due to strong coupling of $Q_i$ 
is not calculable.  Yet noting that only the combination $\lambda S$ 
couples to the strong sector, the contribution to the effective K\"ahler 
potential of $S$ coming from strong coupling physics at the scale 
$\Lambda_s^\prime \approx 4\pi \Lambda_s$ is given by
\begin{equation}
  K = \frac{\Lambda_s^{\prime 2}}{(4\pi)^2}\, 
    {\cal G}\left(\frac{|\lambda S|^2}{\Lambda_s^{\prime 2}}\right),
\label{eq:K-strong}
\end{equation}
where ${\cal G}(x)$ is a polynomial function with the coefficients of 
$O(1)$ up to symmetry factors, and the factor of $4\pi$ is inserted using 
naive dimensional analysis~\cite{Luty:1997fk}.  The quartic correction 
to the K\"ahler potential for $S$ is therefore of $O(\lambda^4/16\pi^2 
\Lambda_s^{\prime 2})$ from the strong sector.

On the other hand, the Coleman--Weinberg potential for $S$ due to 
loops of $M_a$ gives the quartic term of $S$ in the effective K\"ahler 
potential at $O(\lambda^2/\Lambda_s^{\prime 2})$.  Here, we have assumed 
that $\lambda_5$ and $\lambda$ are of the same order of magnitude, and 
$\lambda^2/\Lambda_s^{\prime 2}$ arises from the product of the one-loop 
factor, $1/16\pi^2$, four couplings of $S$, $\lambda^4$, and the inverse 
square of the $M_a$ masses, $1/(\lambda \Lambda_s)^2$.  We thus find 
that for a perturbative value of $\lambda$, i.e. $\lambda \lesssim 4\pi$, 
the calculable correction dominates over the incalculable one in 
Eq.~(\ref{eq:K-strong}).  Indeed, one can show that for $\lambda_5 
\geq \lambda$ the potential has a minimum at $S=0$ with positive 
curvature.\footnote{
  The real parts of $M_a$ become Nambu--Goldstone bosons of a spontaneously 
  broken $SO(6)$ symmetry and hence massless in the limit $\lambda_5 
  \rightarrow \lambda+0$.  Their mass squared goes negative for $\lambda_5 
  < \lambda$, and the new minimum gives a nonvanishing $F$ component 
  for $S_a$, instead of $S$.}
With the renormalized $\lambda_5$, $\lambda$ and $\Lambda_s$ in 
Eq.~(\ref{eq:W_IYIT_2}), the bosons have a mass matrix
\begin{eqnarray}
  \lefteqn{
  m_B^2 = } \nonumber \\ && \left(
    \begin{array}{cccc}
      \! \lambda_5^2 \Lambda_s^2 \!&\! -\lambda \lambda_5 \Lambda_s S \!&\! 
        0 \!&\! 0 \!\\
      \! -\lambda \lambda_5 \Lambda_s S^\dagger \!&\! \lambda_5^2 \Lambda_s^2 
        + \lambda^2 |S|^2 \!&\! 0 \!&\! -\lambda^2 \Lambda_s^2 \!\\
      \! 0 \!&\! 0 \!&\! \lambda_5^2 \Lambda_s^2 \!&\! 
        -\lambda \lambda_5 \Lambda_s S^\dagger \!\\
      \! 0 \!&\! -\lambda^2 \Lambda_s^2 \!&\! -\lambda \lambda_5 \Lambda_s S \!&\! 
        \lambda_5^2 \Lambda_s^2 + \lambda^2 |S|^2 \!
    \end{array} \right),
\nonumber\\
\label{eq:mB2}
\end{eqnarray}
% %
% \begin{eqnarray}
%   \lefteqn{
%   m_B^2 = } \nonumber \\ && \left(
%     \begin{array}{cccc}
%       \lambda_5^2 \Lambda_s^2 & -\lambda \lambda_5 \Lambda_s S & 0 & 0 \\
%       -\lambda \lambda_5 \Lambda_s S^\dagger & \lambda_5^2 \Lambda_s^2 
%         + \lambda^2 |S|^2 & 0 & -\lambda^2 \Lambda_s^2 \\
%       0 & 0 & \lambda_5^2 \Lambda_s^2 & -\lambda \lambda_5 \Lambda_s S^\dagger \\
%       0 & -\lambda^2 \Lambda_s^2 & -\lambda \lambda_5 \Lambda_s S & 
%         \lambda_5^2 \Lambda_s^2 + \lambda^2 |S|^2 
%     \end{array} \right),
% \nonumber\\
% \label{eq:mB2}
% \end{eqnarray}
% %
%
while the fermions
\begin{equation}
  m_F^2 = \left(
    \begin{array}{cc}
      \lambda_5^2 \Lambda_s^2 & -\lambda \lambda_5 \Lambda_s S^\dagger \\
      -\lambda \lambda_5 \Lambda_s S & \lambda_5^2 \Lambda_s^2 + \lambda^2 |S|^2 
    \end{array} \right).
\label{eq:mF2}
\end{equation}
Here, we have used the fact that the kinetic terms for $M_a$ are given by 
$K \approx M_a^\dagger M_a/\Lambda_s^2$.  The curvature $m_S^2$ at the origin, 
defined as $V = V_0 + m_S^2 |S|^2 + O(|S|^4)$, can then be calculated as
\begin{eqnarray}
  \lefteqn{
  m_S^2 =} \nonumber \\ && \frac{5\Lambda_s^2}{32\pi^2} 
    \left( (\lambda_5^2-\lambda^2)^2 
      \ln\frac{\lambda_5^2+\lambda^2}{\lambda_5^2-\lambda^2} 
    + 2 \lambda^2 \lambda_5^2 
      \ln\frac{(\lambda_5^2+\lambda^2)^2}{e \lambda_5^4} \right),
\nonumber\\
\label{eq:mS2}
\end{eqnarray}
and we find that $m_S^2 \geq 0$ for all $\lambda_5 \geq \lambda$.  This 
explicit calculation confirms our power counting, $m_S^2 \approx \lambda^2 
|F_S|^2/16\pi^2 \Lambda_s^2 \approx \lambda^4 \Lambda_s^2/16\pi^2$.

The theory is not calculable for $S \sim \Lambda_s^\prime/\lambda$, and hence 
it in principle allows for a local minimum there~\cite{Intriligator:1996pu}. 
If there is indeed a local minimum at $S \sim \Lambda_s^\prime/\lambda$, 
it also provides a phenomenologically acceptable minimum.  In this paper 
we have picked the minimum close to the origin $S \approx 0$, since we 
know it exists and thus is on a firmer theoretical footing.

To summarize, the complete superpotential of the model is given by the 
combination of Eqs.~(\ref{eq:W}) and (\ref{eq:W_IYIT}):
\begin{equation}
  W = -\lambda_5 S_a M_a - \lambda S_6 M_6 
    + \kappa S_6 f \bar{f} + M f \bar{f}.
\label{eq:complete-IYIT}
\end{equation}
The other possible gauge-invariant, renormalizable terms must be suppressed. 
Their coefficients must be smaller than of $O({\rm min}\{ \Lambda_s, M \})$ 
for dimensionful ones and of $O(1/16\pi^2)$ for dimensionless ones.  The model 
reduces to the one of Eqs.~(\ref{eq:W},~\ref{eq:K}) at low energies.  The 
correspondence of the scales is given by $\mu^2 = \lambda \Lambda_s^2$ 
and $\Lambda \simeq 4\pi \Lambda_s/\lambda$.  Extensions of the model to 
other gauge groups, $Sp(N_c)$ ($N_c > 1$) and $SU(N_c)$ ($N_c > 2$), are 
straightforward.

\section{Theories without Elementary Singlets}
\label{sec:composite}

Models in the previous section contain elementary singlets $S$, so that 
the superpotential terms $S^2$ and $S^3$ must be suppressed ``by hand'' 
to obtain the approximate $U(1)_R$ symmetry in the supersymmetry breaking 
sector.  In this section, we present models that do not contain any 
fundamental singlets.  The effective singlet $S$ arises as a composite 
field at low energies, which allows for natural suppressions of the 
$S^2$ and $S^3$ terms in the low-energy effective superpotentials. 
One class is our previous work~\cite{Murayama:2006yf} and its 
straightforward generalizations based on the supersymmetry breaking 
mechanism of Ref.~\cite{Intriligator:2006dd}, where the negative quartic 
term in the K\"ahler potential originates from loops of light fields. 
However, the success of our scheme is not limited to this class of models. 
We also show other classes of models which enjoy comparable successes, 
with tree-level or dynamical origin of the negative quartic term.

\subsection{Models of Ref.~\cite{Murayama:2006yf} and their 
straightforward variations}
\label{subsec:MN}

We begin by reviewing a class of models constructed in our previous 
work Ref.~\cite{Murayama:2006yf}.  Strictly speaking, these models do 
not reduce to the one given by Eqs.~(\ref{eq:W},~\ref{eq:K}), since 
there are several ``$S$'' fields that carry nonvanishing $F$-component 
expectation values, $F_S$.  This slightly changes the situation.  For 
example, turning on expectation values of the messengers cannot absorb 
all the $F_S$'s, so it does not lead to a supersymmetric minimum. 
Nonetheless, the basic structure of the models is still that of 
Section~\ref{sec:framework}, and many of the analyses there remain 
without any essential changes.  At the qualitative level, even the 
constraint from tunneling can persist.  We simply have to reinterpret 
the tunneling to the supersymmetric minimum as that to a lower, 
phenomenologically unacceptable minimum, which may arise by turning 
on messenger expectation values.

The models employ $SU(N_c)$, $SO(N_c)$ or $Sp(N_c)$ gauge theories with 
massive vector-like quarks.  Here we consider an $SU(N_c)$ gauge theory 
for definiteness, and denote quark and antiquark chiral superfields by 
$Q^i$ and $\bar{Q}^i$ ($i = 1,\cdots,N_f$).  We take the number of quark 
flavors to be in the range $N_c + 1 \leq N_f < \frac{3}{2} N_c$.  The 
tree-level superpotential in this sector is given by
\begin{equation}
  W = m_{ij} \bar{Q}^i Q^j.
\label{eq:W_ISS-1}
\end{equation}
We adopt the basis in which the quark mass matrix is diagonal, $m_{ij} 
= -m_i \delta_{ij}$ with $m_i$ real and positive.  We consider that all 
the masses are different to avoid (potentially) unwanted Nambu--Goldstone 
bosons, and assume that they are ordered as $ m_1 > m_2 > \cdots 
> m_{N_f} > 0$ without loss of generality.

For $m_i \ll \Lambda_s$, the theory breaks supersymmetry on a local minimum, 
where $\Lambda_s$ is the dynamical scale of $SU(N_c)$~\cite{Intriligator:2006dd}. 
After integrating out the excitations of masses of order $(m \Lambda_s)^{1/2}$, 
the relevant degrees of freedom are $S^{ij} = \bar{Q}^i Q^j/\Lambda_s$ 
($i,j = N_f-N_c+1,\cdots,N_f$) with the superpotential
\begin{equation}
  W = - m_i \Lambda_s S^{ii},
\label{eq:W_ISS-2}
\end{equation}
where we have assumed $m_i \sim m$ for simplicity.  These degrees of 
freedom obtain masses of order $(m \Lambda_s)^{1/2}/4\pi$ due to the 
corrections to the K\"ahler potential.  This, therefore, reproduces 
the essential structure of Eqs.~(\ref{eq:W},~\ref{eq:K}).

The complete superpotential in the electric theory is given by the combination 
of the quark mass terms, Eq.~(\ref{eq:W_ISS-1}), and general interactions of 
the quarks with the messengers~\cite{Murayama:2006yf}:
\begin{equation}
  W = m_{ij} \bar{Q}^i Q^j 
    + \frac{\lambda_{ij}}{M_*} \bar{Q}^i Q^j f \bar{f} 
    + M f \bar{f},
\label{eq:complete-MN}
\end{equation}
where $M_*$ is the cutoff scale of the theory, and $\lambda_{ij}$ are 
dimensionless constants.  The correspondence between the scales of the 
present model and those in Section~\ref{sec:framework} is given by $\mu^2 
\simeq m \Lambda_s$, $\kappa \simeq \lambda \Lambda_s/M_*$ and $\Lambda 
\simeq 4\pi (m \Lambda_s)^{1/2}$, where we have assumed $m_i \sim m$ 
and $\lambda_{ij} \sim \lambda$.

We finally comment on an example of straightforward variations of the models 
reviewed above.  In the above models, the effective supersymmetry breaking 
fields are two-body composite states, $S^{ii} \sim \bar{Q}^i Q^i$, so that 
the supersymmetry breaking superpotential of Eq.~(\ref{eq:W_ISS-2}) comes 
from dimension-two operators in the ultraviolet, Eq.~(\ref{eq:W_ISS-1}). 
We can, however, also consider models in which the supersymmetry breaking 
fields are $n$-body composite states with $n>2$.  Consider, for example, 
an $SU(N_c)$ gauge theory with $N_f$ massless vector-like quarks, $Q^i$ and 
$\bar{Q}^i$ ($i = 1,\cdots,N_f$), and a massless adjoint chiral superfield 
$X$.\footnote{
  The absence of the masses is not crucial.  They just have to be suppressed 
  sufficiently so that they do not alter the essential dynamics.}
The superpotential of the theory is then
\begin{equation}
  W = \frac{\lambda}{3} {\rm Tr}X^3 - \zeta_{ij} \bar{Q}^i X Q^j.
\label{eq:W_Kut-1}
\end{equation}
For $\frac{1}{2}N_c+1 \leq N_f < \frac{2}{3}N_c$, this theory has a 
dual magnetic description which is infrared free~\cite{Kutasov:1995ve}. 
The dual theory is an $SU(2N_f-N_c)$ gauge theory with $N_f$ vector-like 
quarks, $q_i$ and $\bar{q}_i$, an adjoint, $Y$, and elementary singlets, 
$M^{ij} = \bar{Q}^i Q^j/\Lambda_s$ and $S^{ij} = \bar{Q}^i X Q^j/\Lambda_s^2$. 
The magnetic theory has the superpotential
\begin{equation}
  W = -\frac{\lambda}{3} {\rm Tr}Y^3 
    + \frac{\lambda}{\Lambda_s} M^{ij} \bar{q}_i Y q_j
    + \lambda S^{ij} \bar{q}_i q_j
    - \zeta_{ij} \Lambda_s^2 S^{ij}.
\label{eq:W_Kut-2}
\end{equation}
(The first term is absent for $N_f = \frac{1}{2}N_c+1$.)  Here, we 
have normalized the fields $q_i$, $\bar{q}_i$, $Y$, $M^{ij}$ and 
$S^{ij}$ to have canonical mass dimensions in the infrared, and we 
have taken $\Lambda_{\rm el} = \Lambda_{\rm mag} \equiv \Lambda_s$ 
for simplicity.\footnote{
  A similar superpotential to Eq.~(\ref{eq:W_Kut-2}) is obtained for 
  $N_f = \frac{1}{2}(N_c+1)$ if $N_c$ is odd.  In this case the relevant 
  infrared degrees of freedom are $q_i$, $\bar{q}_i$ and $S^{ij}$, 
  so that the first two terms of Eq.~(\ref{eq:W_Kut-2}) are absent. 
  The model also works in this case, since a possible nonperturbative 
  superpotential term $({\rm det}S^{ij})^2 S^{-1}_{kl} 
  M^{lk}$~\cite{Csaki:1998fm} is irrelevant.}

We find that the last two terms of Eq.~(\ref{eq:W_Kut-2}) have the 
identical structure with the corresponding terms in the previous 
model.\footnote{
  A more complicated case without an accidental low-energy $U(1)_R$ 
  symmetry was considered in Ref.~\cite{Amariti:2006vk}.}
The $S^{ij}$ fields can thus serve the role of the supersymmetry 
breaking fields.  The stability of $S^{ij}$ is ensured by loops of 
the dual quarks $q_i$ and $\bar{q}_i$, and potentially unwanted light 
fields obtain masses from higher dimension operators omitted in 
Eq.~(\ref{eq:W_Kut-1}).  (Under the existence of higher dimension 
operators, an appropriate vacuum must be chosen in the dual magnetic 
theory.)  Together with the couplings to the messengers
\begin{equation}
  W = \frac{\eta_{ij}}{M_*^2} \bar{Q}^i X Q^j f \bar{f},
\label{eq:W_Kut-3}
\end{equation}
this provides gauge mediation models in which the effective supersymmetry 
breaking fields are three-body composite states ($n=3$ in the language 
of Section~\ref{subsec:S}).  According to the general discussions in 
Section~\ref{subsec:S}, the models require small couplings $\zeta_{ij}$ 
or an enhancement of the operators of Eq.~(\ref{eq:W_Kut-3}).

\subsection{{\boldmath $SO(10)$} model with {\boldmath $\psi(16)$} 
and {\boldmath $H(10)$}}
\label{subsec:SO10}

A general philosophy advocated in Ref.~\cite{Murayama:2006yf} is to discard 
a $U(1)_R$ symmetry altogether at the level of a fundamental theory.  An 
approximate $U(1)_R$ symmetry should then arise in the low-energy effective 
theory as an accidental property of the supersymmetry breaking sector. 
Presumably the earliest calculable model of supersymmetry breaking without 
a $U(1)_R$ symmetry is an $SO(10)$ gauge theory with two chiral superfields, 
$\psi({\bf 16})$ and $H({\bf 10})$~\cite{Murayama:1995ng}.  This theory 
breaks supersymmetry under the existence of an $H$ mass term, and can be 
regarded as a continuous deformation of an incalculable model of supersymmetry 
breaking, $SO(10)$ with a single ${\bf 16}$~\cite{Affleck:1984mf}, since 
they hold the same Witten index~\cite{Witten:1982df}.  We can thus use 
this theory to construct a model of gauge mediation by coupling it to the 
messengers, along the lines of Ref.~\cite{Murayama:2006yf}.

\begin{table}
\centering
\begin{tabular}{|c|cc|cc|} \hline
  & $\psi({\bf 16})$ & $H({\bf 10})$ & $X=\psi\psi H$ & $Y=H^2$ \\ \hline
  $U(1)_R$ & $-3$ & $1$ & $-5$ & $2$\\
  $U(1)_M$ & $-1$ & $2$ & $0$ & $4$\\ \hline
\end{tabular}
\caption{Global symmetries of the $SO(10)$ model in the absence 
 of a superpotential.}
\label{table:U(1)}
\end{table}
In the absence of a superpotential, the theory has global symmetries listed 
in Table~\ref{table:U(1)}.  These symmetries are explicitly broken under the 
existence of the most general renormalizable superpotential consistent with 
the gauge symmetry:
\begin{equation}
  W = \lambda \psi \psi H - \frac{m}{2} H^2.
\label{eq:W_SO10-1}
\end{equation}
The general $D$-flat directions are parameterized by gauge-invariant 
polynomials $X = \psi\psi H$ and $Y = H^2$.  At a generic point in $X$-$Y$ 
space, the gauge group is broken to $SO(7)$,\footnote{
  This is a non-standard embedding, where $SO(7)$ is embedded into the 
  $SO(8)$ subgroup of $SO(10)$ after we use the triality that switches 
  the vector representation and one of the Majorana--Weyl spinor 
  representations.}
whose gaugino condensation generates a nonperturbative superpotential
\begin{equation}
  W_{\rm np} = c \frac{\Lambda_s^{21/5}}{X^{2/5}},
\label{eq:W_SO10-2}
\end{equation}
where $c$ is a calculable $O(1)$ numerical coefficient, and $\Lambda_s$ the 
dynamical scale of $SO(10)$.  Since the value of $c$ is not important in the 
rest of the discussions, we set $c=1$ by suitably changing the normalization 
of $\Lambda_s$.

The model is calculable when $\lambda \ll 1$ and $m \ll \Lambda_s$.  In this 
limit, we can first ignore the mass term $-m H^2/2$ and find a moduli space 
of supersymmetric vacua
\begin{equation}
  \langle X \rangle = \left( \frac{2}{5\lambda} \right)^{5/7} \Lambda_s^3,
\qquad
  Y:\,\, \mbox{arbitrary}.
\label{eq:moduli_SO10}
\end{equation}
One can verify that the $U(1)_M$ anomalies are saturated by the composite 
$Y$ alone.  As long as $\lambda \ll 1$ and hence $\langle X \rangle \gg 
\Lambda_s^3$ the theory is weakly coupled, and the K\"ahler potential for 
$X$ and $Y$ can be worked out with the tree-level approximation.  We use 
a similar technique to that in~\cite{Bagger:1994hh}.  The result is
\begin{equation}
  K = x^2 + \frac{1}{\sqrt{2} x} |X| + \frac{1}{4x^2} |Y|^2.
\label{eq:K_SO10}
\end{equation}
Here, $x$ is the real positive solution to the equation $\partial 
K/\partial x = 0$:
\begin{equation}
  4 x^4 - \sqrt{2}|X| x - |Y|^2 = 0,
\label{eq:x}
\end{equation}
which can be solved analytically using Ferrari's method.  With $X$ integrated 
out along the moduli space of Eq.~(\ref{eq:moduli_SO10}), the low-energy 
theory is one with $Y$ alone, whose K\"ahler potential can be expanded 
around the origin as
\begin{equation}
  K = \frac{3}{2} |\langle X \rangle|^{2/3} 
    + \frac{|Y|^2}{2|\langle X \rangle|^{2/3}} 
    - \frac{|Y|^4}{6|\langle X \rangle|^2} + O(|Y|^6).
\label{eq:Kahler-Y}
\end{equation}
It is guaranteed that this K\"ahler potential depends only on the combination 
$|Y|^2$ because of the $U(1)_M$ invariance of the theory in the absence 
of the mass term $-m Y/2$.  We thus find that the low-energy theory, 
characterized by the K\"ahler potential of Eq.~(\ref{eq:Kahler-Y}) and 
the linear superpotential term $W = -m Y/2$, has an {\it accidental}\/ 
$U(1)_R$ symmetry, under which $Y$ carries a charge of $+2$.

The rest of the discussion reduces to the general one in 
Section~\ref{sec:framework}.  Note that the negative coefficient for
the quartic term in the K\"ahler potential originates not from one-loop 
effects as in the models in Section~\ref{subsec:MN} but rather from the 
tree-level K\"ahler potential along the $D$-flat directions.  Correspondingly, 
there are no other light fields in the theory to generate the quartic 
term, which makes the model more easily compatible with cosmology. 
The coupling to the messengers is given by
\begin{equation}
  W = \frac{\eta}{2 M_*} H^2 f \bar{f}.
\label{W_SO10-3}
\end{equation}
The correspondence of the scales can be worked out easily by 
canonically normalizing the $Y$ field in Eq.~(\ref{eq:Kahler-Y}): 
$S \equiv Y/\sqrt{2} |\langle X \rangle|^{1/3}$.  It is given by 
$\mu^2 \simeq m \Lambda_s/\lambda^{5/21}$, $\kappa \simeq \eta 
\Lambda_s/\lambda^{5/21} M_*$ and $\Lambda \simeq \Lambda_s/\lambda^{5/21}$. 
For an appropriate range of the parameters, the superpotential of the 
model can be a generic one compatible with the gauge symmetry, as in 
the models of Ref.~\cite{Murayama:2006yf}.

In fact, the basic dynamics of the model just described is more general. 
Consider a model of dynamical supersymmetry breaking in which some of 
the classical flat directions are lifted by superpotential interactions. 
By choosing these interactions appropriately, one can make expectation 
values of fields larger than the dynamical scale, and thus make the 
model calculable.  Now, if the model allows for making only one field 
$S$ significantly lighter than the rest of the excitations (such as the 
$Y$ field above), then one can write a low-energy effective theory that 
contains only a single composite field $S$.  By shifting the origin of 
$S$ such that $\langle S \rangle = 0$ at the minimum, the superpotential 
contains a linear term, and the K\"ahler potential takes generically 
the form of Eq.~(\ref{eq:K}).  We can then construct a model of gauge 
mediation simply by coupling the gauge invariant operator $S$ to the 
messenger bilinear $f\bar{f}$ in the superpotential.

A small variation of this picture is obtained, for example, in the $SU(5)$ 
model with $A({\bf 10})$, $F({\bf 5})$, and two $\bar{F}_i({\bf 5}^*)$ 
($i=1,2$)~\cite{Murayama:1995ng}.  This model can be viewed as a 
continuous deformation of the incalculable model with only $A$ and one 
$\bar{F}$~\cite{Affleck:1983vc}, once a mass term is given to a pair of 
$F$ and $\bar{F}$.  The most general superpotential of the model is
\begin{equation}
  W = \lambda A \bar{F}_1 \bar{F}_2 + \lambda' A A F - m \bar{F}_1 F,
\label{eq:calc-SU5-tree}
\end{equation}
while the nonperturbative superpotential is
\begin{equation}
  W_{\rm np} = \frac{\Lambda_s^6}{[(A \bar{F}_1 \bar{F}_2) (A A F)]^{1/2}},
\label{eq:calc-SU5-np}
\end{equation}
where $\Lambda_s$ is the dynamical scale of $SU(5)$.  There is no $U(1)_R$ 
symmetry, but there is a global $SU(2) \times U(1)$ symmetry in this model. 
In terms of the gauge-invariant polynomials $X = (1/2)(A \bar{F}_1 \bar{F}_2)$, 
$Y = (1/\sqrt{2}) (A A F)$ and $S_i = \bar{F}_i F$, the tree-level K\"ahler 
potential can be worked out and expanded in $S_i$ as
\begin{eqnarray}
  K &=& 6(|X| + |Y|)^{2/3}
    + \frac{\sum_i S_i^\dagger S_i}{2(|X| + |Y|)^{2/3}}
\nonumber\\
  && - \frac{(\sum_i S_i^\dagger S_i)^2}{24(|X| + |Y|)^{2}} + O(S_i^6).
\label{eq:calc-SU5-K}
\end{eqnarray}
The global $SU(2) \times U(1)$ invariance of the theory guarantees that 
it depends on $S_i$ only through the combination $\sum_i S_i^\dagger S_i$. 
After minimizing the superpotential without the mass term ($m = 0$), both 
$X$ and $Y$ are fixed and can be integrated out.  The low-energy theory 
consists of $S_i$ alone, which saturates the $SU(2)\times U(1)$ anomalies. 
Given the negative quartic term in the K\"ahler potential, the mass term 
breaks supersymmetry with a stable minimum at the origin $S_i = 0$.  One 
can verify that both $S_1$ and $S_2$ acquire positive squared masses. 
In fact, this model generalizes to $SU(2k+1)$ with an antisymmetric 
tensor $A$, one fundamental $F$ and $(2k-2)$ antifundamentals $\bar{F}_i$. 
With $A^k F$ and $A \bar{F}_i \bar{F}_j$ terms in the superpotential 
and the nonperturbative superpotential $W_{\rm np} \propto [(A^k F) 
{\rm Pf}(A\bar{F}_i \bar{F}_j)]^{-1/2}$, the low-energy theory is given 
in terms of $S_i = \bar{F}_i F$ that match the anomalies of the global 
$Sp(k-1) \times U(1)$ symmetry.  A mass term $m \bar{F}_1 F$ would break 
supersymmetry with a stable minimum at the origin.  Then the coupling 
to the messengers $\bar{F}_i F f \bar{f}/M_*$ makes gauge mediation 
possible.

\subsection{Models with incalculable K\"ahler potentials}
\label{subsec:SU2}

The first model we present here uses the supersymmetry breaking theory 
of Ref.~\cite{Intriligator:1994rx}, based on the phenomenon of quantum 
smooth-out of classical singularities in moduli space.  Consider an 
$SU(2)$ gauge theory with a single chiral superfield $Q$ in the $I=3/2$ 
representation.  The gauge invariant chiral operator in this theory is 
$u = QQQQ$, and we introduce the following tree-level superpotential:
\begin{equation}
  W = -\frac{\zeta}{M_*} u,
\label{eq:W_SU2}
\end{equation}
where $M_*$ is the cutoff scale of the theory, presumably of order 
$M_{\rm Pl}$, and $\zeta$ a dimensionless constant.  Since $u$ saturates 
nontrivial 't~Hooft anomaly matching conditions~\cite{'tHooft:AMC}, we 
expect that $u$ is the only low-energy degree of freedom.  The K\"ahler 
potential for $u$ is then given by
\begin{equation}
  K = \Lambda_s^2\, {\cal G}\left(\frac{|u|^2}{\Lambda_s^8}\right),
\label{eq:K_SU2}
\end{equation}
where $\Lambda_s$ is the dynamical scale of $SU(2)$ gauge interactions. 
For $|u| \ll \Lambda_s^4$, ${\cal G}(x)$ is expected to be a polynomial 
function with the coefficients of $O(1)$ up to symmetry factors --- 
the classical singularity at the origin of $u$ is smoothed out by 
quantum effects.

Denoting the field with the canonical dimension by $S = u/\Lambda_s^3$, 
the low-energy super- and K\"ahler potentials for $|S| \ll \Lambda_s$ take 
the form given by the first term of Eq.~(\ref{eq:W}) and Eq.~(\ref{eq:K}), 
respectively.  In the present theory, however, the sign of the quartic 
term in Eq.~(\ref{eq:K}) is incalculable, while it must be negative in 
order for the model to work.  We thus make a dynamical assumption that 
the sign of this term is negative.

The complete superpotential of the model is given by
\begin{equation}
  W = -\frac{\zeta}{M_*} u + \frac{\eta}{M_*^3} u f \bar{f} + M f \bar{f},
\label{eq:complte-SU2}
\end{equation}
where $\eta$ is a dimensionless constant.  The model reduces at low 
energies to that of Eqs.~(\ref{eq:W},~\ref{eq:K}), with $\mu^2 = \zeta 
\Lambda_s^3/M_*$, $\kappa = \eta \Lambda_s^3/M_*^3$ and $\Lambda \simeq 
\Lambda_s$.  In addition to the terms in Eq.~(\ref{eq:complte-SU2}), the 
most general terms consistent with the gauge symmetry may present.  The 
coefficients of these terms need not be suppressed, since the constraints 
of Eq.~(\ref{eq:bound-MS-kS}) are almost automatically satisfied because 
of the compositeness of $S$.

As discussed in Section~\ref{subsec:S}, the dimensionless coupling $\zeta$ 
must be small in order for the model to work (for $\eta \sim 1$ and $M_* 
\sim M_{\rm Pl}$; see Eq.~(\ref{eq:gen-bound}).)  Successful parameter 
regions include, for example, $\Lambda_s \simeq 10^{15}~{\rm GeV}$, $M_* 
\simeq 10^{18}~{\rm GeV}$, $\zeta \simeq 10^{-8}$, $\eta \simeq 1$, 
and $M \simeq 10^6~{\rm GeV}$.

A nearly identical analysis can be made on an $SU(6)$ gauge theory 
with a rank-three antisymmetric tensor $A^{ijk}$.  For general $D$-flat 
configurations, the gauge group is broken to $SU(3) \times SU(3)$, 
each of which develops a gaugino condensate.  Depending on the relative 
phase between the two condensates, the nonperturbative superpotential
\begin{equation}
  W_{\rm np} = \bigl( 1+(-1)^{1/3} \bigr) \frac{\Lambda_s^5}{(A^4)^{1/2}},
\end{equation}
can identically vanish.\footnote{
  For an alternative derivation of inequivalent branches, 
  see~\cite{Csaki:1996zb}.}
The composite field $A^4$ saturates the $U(1)_R$ and $U(1)_R^3$ anomalies, 
so we expect it to have a nonsingular K\"ahler potential at the origin. 
An introduction of a linear term in $A^4$ would then break supersymmetry. 
As in the $SU(2)$ model, however, the quartic term in the K\"ahler potential 
is not calculable.  We thus have to assume that its coefficient is negative 
in order to use this theory.

Yet another example is $SO(N)$ theories with $N-4$ vectors.  They have 
two inequivalent branches, one with and the other without a dynamical 
superpotential~\cite{Intriligator:1995id}.  All anomalies are saturated 
by the mesons $M^{ij} = Q^i Q^j$.\footnote{
  Discrete anomalies are not matched because of a condensate 
  $\langle Q^{N-4} {\cal W}^\alpha {\cal W}_\alpha \rangle \neq 0$; 
  see~\cite{Csaki:1997aw}.}
Adding a mass term to just one of the flavors, the theory breaks 
supersymmetry.  Again the quartic term in the K\"ahler potential 
is not calculable and we have to simply assume that its coefficient 
is negative to use this theory.

\section{Related Models}
\label{sec:related}

In this section we present models that do not exactly fall in the category 
discussed in Section~\ref{sec:framework}.  We first present models in which 
the low-energy effective theories contain more than one field, $S$.  In 
general, these theories have multiple composite fields $X_i$ ($i=1,2,\cdots$) 
at low energies, which are stabilized due to complicated K\"ahler potentials. 
Models of gauge mediation are then obtained by coupling the degree of freedom 
responsible for supersymmetry breaking to the messengers in the superpotential. 
In the case that the K\"ahler potentials are complicated, the low-energy 
fields $X_i$ cannot be regarded simply as multiple copies of an $S$ field, 
in contrast with the case in some of the previous models such as the 
ones in Section~\ref{subsec:MN}.

We then consider models that do not contain any degree of freedom which 
is significantly lighter than the dynamical scale.  While these models are 
generally incalculable, models of gauge mediation can be obtained by coupling 
the messengers to appropriate composite operators.

While the models discussed in this section do not have an identical 
low-energy structure to those of Section~\ref{sec:framework}, they share 
many features.  In particular, the basic constructions of the models are 
quite similar --- we simply prepare a supersymmetry breaking model that 
has a stable supersymmetry breaking minimum (either global or local), and 
then couple the operator responsible for supersymmetry breaking to the 
bilinear of (generically massive) messengers.  Many of the general analyses 
in Section~\ref{sec:framework} also persist.  In particular, a general 
constraint on parameters in Eq.~(\ref{eq:gen-bound}) persists, despite 
the fact that the powers of $\zeta$ appearing in the gauge-mediated and 
gravity-mediated contributions in Eqs.~(\ref{eq:m_SUSY-2},~\ref{eq:m32-2}) 
can now be different.  Here, $\zeta$ represents the coefficient in front 
of the operator responsible for supersymmetry breaking in the superpotential. 
We will prove this fact in Section~\ref{subsec:multiple}.

\subsection{Models with multiple low-energy fields}
\label{subsec:multiple}

A class of theories that breaks supersymmetry dynamically is chiral gauge 
theories which do not possess classical flat directions, and in which global 
symmetries are spontaneously broken~\cite{Affleck:1983vc,Affleck:1984uz,%
Affleck:1984xz}.  These theories have stable supersymmetry breaking vacua, 
and one can construct models of gauge mediation by coupling these theories 
to (generically massive) messengers $f, \bar{f}$.  Here we present one such 
theory explicitly, and analyze its relations to the class of models 
discussed in Section~\ref{sec:framework}.

Consider an $SU(3) \times SU(2)$ gauge theory with the matter content 
$Q({\bf 3},{\bf 2})$, $U({\bf 3}^*,{\bf 1})$, $D({\bf 3}^*,{\bf 1})$ and 
$L({\bf 1},{\bf 2})$, with the tree-level superpotential $W = \zeta QDL$. 
This theory breaks supersymmetry at the vacuum with expectation values 
for the fields $v \sim \Lambda_s/\zeta^{1/7}$~\cite{Affleck:1984xz}.  Here, 
$\Lambda_s$ is the dynamical scale of $SU(3)$, which we assume to be larger 
than that of $SU(2)$.  The vacuum energy is given by $V \sim \zeta^{10/7} 
\Lambda_s^4$.

It is useful to analyze the theory in terms of the gauge-invariant composite 
fields: $X_1 = QDL$, $X_2 = QUL$ and $X_3 = {\rm det}(\bar{Q}_i Q^j)$, 
where $\bar{Q}_i \equiv (D, U)$ and $j=1,2$ is the $SU(2)$ index.  Including 
nonperturbative effects, the low-energy effective superpotential is
\begin{equation}
  W = \zeta X_1 + 2\frac{\Lambda_s^7}{X_3}.
\label{eq:W_32}
\end{equation}
For $\zeta \ll 1$, expectation values for the fields are much larger than 
the dynamical scale, $v \gg \Lambda_s$, so that the K\"ahler potential 
is well approximated by the tree-level one.  In terms of the composite 
fields, it is given by~\cite{Bagger:1994hh}
\begin{equation}
  K = 24\frac{A+B x}{x^2},
\label{eq:K_32}
\end{equation}
where $A = (X_1^\dagger X_1 + X_2^\dagger X_2)/2$, $B = (X_3^\dagger 
X_3)^{1/2}/3$, and
\begin{equation}
  x = 4\sqrt{B}\, \cos\left(\frac{1}{3}\arccos\frac{A}{B^{3/2}}\right).
\label{eq:x_32}
\end{equation}
By minimizing the resulting scalar potential, we find that the minimum 
is at
\begin{equation}
  \langle X_1 \rangle \simeq 0.50 \frac{\Lambda_s^3}{\zeta^{3/7}},
\quad
  \langle X_2 \rangle = 0,
\quad
  \langle X_3 \rangle \simeq 2.58 \frac{\Lambda_s^4}{\zeta^{4/7}},
\label{eq:32_VEV}
\end{equation}
with
\begin{equation}
  F_{X_1} \simeq -2.57\, \zeta^{3/7} \Lambda_s^4,
\quad
  F_{X_2} = 0,
\quad
  F_{X_3} \simeq 3.42\, \zeta^{2/7} \Lambda_s^5,
\label{eq:32_F-VEV}
\end{equation}
where $F_{X_i}$ represent the vacuum expectation values for the auxiliary 
components of chiral superfields $X_i$.  We can thus construct a model of 
gauge mediation by coupling $X_1$ to the messengers in the superpotential.

The relevant superpotential for the messengers is
\begin{equation}
  W = \frac{\eta}{M_*^2} QDL f \bar{f} + M f \bar{f},
\label{eq:W_32-mess}
\end{equation}
where $M_*$ is the cutoff scale of the theory.  The supersymmetric and 
holomorphic supersymmetry breaking masses for the messengers are given by
\begin{equation}
  M_{\rm mess} = M + \frac{\eta}{M_*^2} \langle X_1 \rangle
    \approx M + \frac{\eta \Lambda_s^3}{\zeta^{3/7} M_*^2},
\label{eq:32-M_mess}
\end{equation}
and
\begin{equation}
  F_{\rm mess} = \frac{\eta}{M_*^2} \langle F_{X_1} \rangle 
    \approx \frac{\eta \zeta^{3/7} \Lambda_s^4}{M_*^2},
\label{eq:32-F_mess}
\end{equation}
where we have omitted $O(1)$ coefficients.  The condition for avoiding 
tachyonic messengers is
\begin{equation}
  M_{\rm mess} \gtrsim \frac{\eta^{1/2} \zeta^{3/14} \Lambda_s^2}{M_*}.
\label{eq:32-cond}
\end{equation}
As long as this condition is satisfied, the minimum in the original theory 
stays as a local supersymmetry-breaking minimum in the theory with the 
messengers.

The resulting gauge-mediated contribution to the scalar and gaugino masses 
in the SSM sector is of order
\begin{equation}
  m_{\rm SUSY} \approx \frac{g^2}{16\pi^2} 
    \frac{\zeta^{3/7} \eta \Lambda_s^4}{M_*^2 M_{\rm mess}},
\label{eq:32-GMSB}
\end{equation}
while generic gravity-mediated contributions to the SSM-sector scalars, 
arising from K\"ahler potential terms of the form $Q^\dagger Q \Phi^\dagger 
\Phi/M_{\rm Pl}^2$, $U^\dagger U \Phi^\dagger \Phi/M_{\rm Pl}^2$ and so on, 
are of order
\begin{equation}
  m_{3/2} \approx \frac{\zeta^{5/7} \Lambda_s^2}{M_{\rm Pl}},
\label{eq:32-grav}
\end{equation}
where $\Phi$ represents matter and Higgs superfields in the SSM sector. 
To obtain $m_{\rm SUSY} = O(100~{\rm GeV}\!\sim\!1~{\rm TeV})$, we need 
to take
\begin{equation}
  \frac{\zeta^{3/7} \eta \Lambda_s^4}{M_*^2 M_{\rm mess}} 
    \approx 100~{\rm TeV}.
\label{eq:32-mess-scale}
\end{equation}
The tunneling rate from the supersymmetry breaking minimum to the true 
supersymmetric (runaway) minimum does not give a very strong constraint 
on the parameters.

\ From the expressions in Eqs.~(\ref{eq:32-GMSB},~\ref{eq:32-grav}), we 
find that the ratio of gravity- to gauge-mediated contributions is given by
\begin{eqnarray}
  \frac{m_{3/2}}{m_{\rm SUSY}}
    &\approx& \frac{16\pi^2}{g^2} \frac{\zeta^{2/7} M_*^2 M_{\rm mess}}
      {\eta \Lambda_s^2 M_{\rm Pl}}
\nonumber\\
    &\gtrsim& 100\, \frac{\zeta^{1/2} M_*}{\eta^{1/2} M_{\rm Pl}}.
\label{eq:32-ratio}
\end{eqnarray}
Note that this inequality is identical to the corresponding one in 
Section~\ref{subsec:S}, Eq.~(\ref{eq:ratio}), despite the fact that the powers 
of $\zeta$ in Eqs.~(\ref{eq:32-cond},~\ref{eq:32-GMSB},~\ref{eq:32-grav}) are 
different from the corresponding ones in Section~\ref{subsec:S}.  Therefore, 
the general bound in Eq.~(\ref{eq:gen-bound}) also applies to the present 
case.  In particular, the coupling $\zeta$ must be suppressed for $\eta 
\sim 1$ and $M_* \sim M_{\rm Pl}$.  An example of phenomenologically 
successful parameter regions in the present model is $\zeta \simeq 10^{-8}$, 
$\eta \simeq 1$, $\Lambda_s \simeq 10^{12-13}~{\rm GeV}$, $M_* \simeq 
10^{18}~{\rm GeV}$ and $M \simeq 10^{5-6}~{\rm GeV}$.

One can show that the general bound of Eq.~(\ref{eq:gen-bound}) applies quite 
generally in the present class of theories.  Suppose that the gauge-invariant 
chiral superfield operator ${\cal O}$ that couples to $f\bar{f}$ and 
gives $F_{\rm mess}$ and (a part of) $M_{\rm mess}$ consists of $n$ elementary 
fields, ${\cal O} \sim Q^n$, and the superpotential term generating the 
auxiliary component expectation value for ${\cal O}$ is given by $W = \zeta 
{\cal O}/M_*^{n-3}$.  (The $SU(3) \times SU(2)$ model discussed above 
corresponds to the case with $n=3$.)  Characteristic field expectation 
values, $\langle Q \rangle$, depend on the nonperturbatively generated 
superpotential, but can in general be parameterized by $\langle Q \rangle 
\sim \Lambda_s/\zeta^\alpha$, where $\Lambda_s$ is the dynamical scale for 
the relevant gauge interactions and $\alpha > 0$.  (The $SU(3) \times SU(2)$ 
model has $\alpha = 1/7$).  The $F$-term expectation value for $Q$ is 
then given by $F_Q \sim \zeta \langle Q \rangle^{n-1}/M_*^{n-3} \sim 
\zeta^{1-(n-1)\alpha} \Lambda_s^{n-1}/M_*^{n-3}$ (which implies $\alpha 
< 1/(n-1)$ since $F_Q$ should vanish for $\zeta \rightarrow 0$).  The 
coupling to the messengers takes the form $W = (\eta/M_*^{n-1}) {\cal O} 
f\bar{f}$, so the messenger masses are given by
\begin{eqnarray}
  M_{\rm mess} &\approx& 
    M + \frac{\eta \Lambda_s^n}{\zeta^{n\alpha} M_*^{n-1}},
\label{eq:32-gen-Mmess} \\
  F_{\rm mess} &\approx& 
    \frac{\zeta^{1-2(n-1)\alpha} \eta \Lambda_s^{2n-2}}{M_*^{2n-4}},
\label{eq:32-gen-Fmess}
\end{eqnarray}
where we have used the fact that the K\"ahler potential is approximately 
canonical in terms of the ultraviolet fields $Q$ for $\zeta \lesssim 1$. 
The gauge-mediated and gravity-mediated contributions are given by
\begin{eqnarray}
  m_{\rm SUSY} &\approx& \frac{g^2}{16\pi^2} 
    \frac{\zeta^{1-2(n-1)\alpha} \eta \Lambda_s^{2n-2}}{M_*^{2n-4} M_{\rm mess}},
\label{eq:32-gen-GMSB} \\
  m_{3/2} &\approx& 
    \frac{\zeta^{1-(n-1)\alpha} \Lambda_s^{n-1}}{M_*^{n-3} M_{\rm Pl}},
\label{eq:32-gen-grav}
\end{eqnarray}
respectively.  Using the stability condition for the messengers 
$M_{\rm mess} \gtrsim F_{\rm mess}^{1/2}$, we then find the same 
inequality as Eq.~(\ref{eq:32-ratio}):
\begin{eqnarray}
  \frac{m_{3/2}}{m_{\rm SUSY}}
    &\approx& \frac{16\pi^2}{g^2} 
      \frac{\zeta^{(n-1)\alpha} M_*^{n-1} M_{\rm mess}}
      {\eta \Lambda_s^{n-1} M_{\rm Pl}}
\nonumber\\
    &\gtrsim& 100\, \frac{\zeta^{1/2} M_*}{\eta^{1/2} M_{\rm Pl}},
\label{eq:app-gen-ratio}
\end{eqnarray}
which implies the general bound of Eq.~(\ref{eq:gen-bound}).

\subsection{Incalculable models}
\label{subsec:incalculable}

Consider an $SU(5)$ gauge theory with $\psi({\bf 10})$ and $\phi({\bf 5}^*)$.
This theory does not have a classical flat direction, and has an exact global 
$U(1)_R$ symmetry with the charge assignment $R(\psi) = 1$ and $R(\phi) = -9$. 
The difficulty of satisfying the 't~Hooft anomaly matching condition suggests 
that the $U(1)_R$ symmetry is dynamically broken.  The other exact global 
symmetry of the theory, $U(1)_A$ with the charge assignment $A(\psi) = 1$ 
and $A(\phi) = -3$, may or may not be broken.\footnote{
  The $U(1)_A$ symmetry is most likely not broken as its anomalies can 
  be matched by a simple three-fermion composite $(\psi^{ij} \phi_i) \phi_j$, 
  where $i,j$ are $SU(5)$ fundamental indices and the parenthesis represents 
  the contraction of spinor indices.}
The broken $U(1)_R$ symmetry, together with the absence of a classical flat 
direction, then suggests that supersymmetry is dynamically broken at the 
vacuum~\cite{Affleck:1983vc}.

It is natural to consider that the $U(1)_R$ symmetry and supersymmetry 
are broken, respectively, by the vacuum expectation values of the 
lowest and highest components of the operator $S \equiv {\cal W}^\alpha 
{\cal W}_\alpha$, where ${\cal W}_\alpha$ is the gauge field-strength 
superfield for $SU(5)$.\footnote{
  There are other natural candidate chiral superfields for nonvanishing 
  highest-component expectation values: $\bar{\cal D}^2({\rm Tr} e^{V^T} 
  \psi^\dagger e^V \psi)$ and $\bar{\cal D}^2 (\phi^\dagger e^{-V^T} \phi)$. 
  They are, however, identical to ${\cal W}^\alpha {\cal W}_\alpha$ because 
  of the Konishi anomaly~\cite{Konishi:1983hf}.}
We assume here that this is indeed the case: $\langle S \rangle \sim 
\Lambda_s^3$ and $F_S \sim \Lambda_s^4$, where $\Lambda_s$ is the 
dynamical scale of $SU(5)$.  A model of gauge mediation is then obtained 
by introducing the following Lagrangian terms for the messengers:
\begin{equation}
  {\cal L} = \int\!d^2\theta\, 
    \left( \frac{1}{m^2} f \bar{f} {\cal W}^\alpha {\cal W}_\alpha 
    + M f \bar{f} \right) + {\rm h.c.},
\label{eq:L_SU5}
\end{equation}
where $m$ is a scale associated with the generation of the $f \bar{f} 
{\cal W}^\alpha {\cal W}_\alpha$ term (assuming $O(1)$ dimensionless 
couplings).  The supersymmetric and holomorphic supersymmetry breaking 
masses for the messengers are then given by
\begin{eqnarray}
  M_{\rm mess} &\approx& M + \frac{\Lambda_s^3}{m^2},
\label{eq:SU5-M_mess} \\
  F_{\rm mess} &\approx& \frac{\Lambda_s^4}{m^2}.
\label{eq:SU5-F_mess}
\end{eqnarray}
The gauge-mediated and gravity-mediated contributions are given by
\begin{eqnarray}
  m_{\rm SUSY} &\approx& \frac{g^2}{16\pi^2} 
    \frac{\Lambda_s^4}{m^2 M_{\rm mess}},
\label{eq:SU5-GMSB} \\
  m_{3/2} &\approx& \frac{\Lambda_s^2}{M_{\rm Pl}}.
\label{eq:SU5-grav}
\end{eqnarray}

There are several constraints for the model to work.  First, the messenger 
stability requires that $M_{\rm mess}^2 > F_{\rm mess}$.  Assuming $m \gg 
\Lambda_s$, this implies that $M_{\rm mess}$ is dominated by the first term 
in Eq.~(\ref{eq:SU5-M_mess}) and that
\begin{equation}
  M \gtrsim \frac{\Lambda_s^2}{m}.
\label{eq:SU5_constr-1}
\end{equation}
We then requires the gauge-mediated contribution to dominate the 
gravity-dominated one, leading to
\begin{equation}
  \frac{m^2 M}{\Lambda_s^2 M_{\rm Pl}} \lesssim O(10^{-4}\!\sim\!10^{-3}).
\label{eq:SU5_constr-2}
\end{equation}
Equations~(\ref{eq:SU5_constr-1},~\ref{eq:SU5_constr-2}) imply that the parameter 
$m$ should be at least 3 or 4 orders of magnitude smaller than $M_{\rm Pl}$. 
The first term of Eq.~(\ref{eq:L_SU5}) with such a small $m$ can be easily 
generated, for example, by introducing a pair of vector-like fields $F({\bf r}) 
+ \bar{F}({\bf r}^*)$ under $SU(5)$, as well as the superpotential
\begin{equation}
  W = \frac{\lambda}{M_*} F \bar{F} f \bar{f} + M_F F \bar{F},
\label{eq:W_SU5}
\end{equation}
where $M_*$ is the cutoff scale of the theory, presumably of order $M_{\rm Pl}$, 
and $M_F$ takes a value in the range
\begin{equation}
  \Lambda_s \lesssim M_F \lesssim M_*.
\label{eq:range-M_F}
\end{equation}
After integrating out the $F,\bar{F}$ fields, we obtain the first term of 
Eq.~(\ref{eq:L_SU5}) with
\begin{equation}
  m^2 \approx 8\pi^2 \frac{M_F M_*}{\lambda}.
\label{eq:SU5-m}
\end{equation}
Finally, to obtain $m_{\rm SUSY} = O(100~{\rm GeV}\!\sim\!1~{\rm TeV})$, 
we need
\begin{equation}
  \frac{\Lambda_s^4}{m^2 M} \approx 100~{\rm TeV}.
\label{eq:SU5-mess-scale}
\end{equation}

To summarize, the model has an $SU(5)$ gauge symmetry with the matter content 
$\psi({\bf 10})$, $\phi({\bf 5}^*)$, $F({\bf r})$ and $\bar{F}({\bf r}^*)$. 
The superpotential is given by
\begin{equation}
  W = M_F F \bar{F} + \frac{\lambda}{M_*} F \bar{F} f \bar{f} + M f \bar{f},
\label{eq:complete-SU5}
\end{equation}
together with the other terms compatible with the gauge symmetry.\footnote{
  The fields $F({\bf r})$ and $\bar{F}({\bf r}^*)$ can be replaced by 
  a single field $F({\bf r})$, and $F \bar{F}$ by $F^2/2$, if ${\bf r}$ 
  is a real representation of $SU(5)$.}
For ${\bf r} = {\bf 5}$ or ${\bf 5}^*$, this model can be regarded 
as a continuous deformation of the model discussed at the end of 
Section~\ref{subsec:SO10}.  An example of successful parameter values 
is $\Lambda_s \simeq 10^{9.5}~{\rm GeV}$, $M_* \simeq 10^{18}~{\rm GeV}$, 
$\lambda \simeq 1$, $M_F \simeq 10^{10}~{\rm GeV}$, and $M \simeq 
10^5~{\rm GeV}$.

\section{Generating Small Parameters}
\label{sec:retro}

The models discussed so far have used small parameters, e.g. small 
dimensionless couplings and/or mass parameters that are hierarchically 
smaller than the cutoff scale.  It is rather easy to generate these 
small parameters dynamically, using supersymmetric dynamics.

As an example, let us consider an $SU(2)$ gauge theory with 4 quark 
chiral superfields ${\cal Q}_i$ ($i=1,\cdots,4$).  There are six 
gauge-invariant meson operators constructed out of ${\cal Q}_i$, which 
can be decomposed into a 5-plet, $({\cal Q} {\cal Q})_m$ ($m=1,\cdots,5$), 
and a singlet, $({\cal Q} {\cal Q})$, under the $SP(4) \simeq SO(5)$ 
subgroup of the flavor $SU(4) \simeq SO(6)$ symmetry.  Nonperturbative 
$SU(2)$ dynamics induce vacuum expectation values for these operators 
$({\cal Q} {\cal Q})_m^2 + ({\cal Q} {\cal Q})^2 = \Lambda'^4$, where 
$\Lambda'$ is the dynamical scale of $SU(2)$~\cite{Seiberg:1994bz}. 
We now introduce the superpotential term $W = k Z_m ({\cal Q} {\cal Q})_m$, 
where $Z_m$ is an $SU(2)$-singlet chiral superfield and $k$ a coupling 
constant.  This leads to $\langle ({\cal Q} {\cal Q})_m \rangle = 0$ and 
$\langle ({\cal Q} {\cal Q}) \rangle = \Lambda'^2$, which can be used as 
a general scale generation mechanism through the $({\cal Q} {\cal Q})$ 
operator~\cite{Izawa:1997gs}.  Specifically, we can generate small 
mass and/or dimensionless parameters, simply by replacing them by (some 
powers of) the operator $({\cal Q} {\cal Q})$ suppressed by appropriate 
powers of the cutoff scale.  For a sufficiently large value of $k$, 
this does not disturb the original dynamics of the models.

Another way of generating small parameters is to use the gaugino 
condensation of a supersymmetric pure Yang-Mills theory.  Suppose we 
replace small parameters by (some powers of) the bilinear of the gauge 
field-strength superfield, ${\cal W}^\alpha {\cal W}_\alpha$ suppressed 
by appropriate powers of the cutoff scale.  The low-energy effective 
Lagrangian is then obtained essentially by setting ${\cal W}^\alpha 
{\cal W}_\alpha = \Lambda'^3$, where $\Lambda'$ is the dynamical scale 
of the Yang-Mills theory.  This, therefore, dynamically generates small 
parameters~\cite{Dine:2006gm}.  While this process also generates other 
(small) terms in the superpotential, minima of the potential in the 
original models are maintained in general, with only small shifts 
in expectation values of the fields.

It is model dependent if these scale generation mechanisms lead 
to a theory in which the interactions are the most general ones 
consistent with symmetries.  For the models in our previous work 
Ref.~\cite{Murayama:2006yf}, this question has been discussed 
in Ref.~\cite{Aharony:2006my}.  In the $SO(10)$ model of 
Section~\ref{subsec:SO10}, we can consider a discrete $R$ symmetry 
under which the mass term of the $H$ field is replaced by (some powers 
of) the gaugino condensation of a pure Yang-Mills theory.  For example, 
we can consider a $Z_{7,R}$ symmetry with the charge assignment 
$R(\psi) = -2$ and $R(H) = -1$.  This leads to an unsuppressed 
coupling $\lambda$, with the $H$ mass term given by $m \approx 
({\cal W}^\alpha {\cal W}_\alpha)^2/M_*^5 = \Lambda'^6/M_*^5$.

\section{Conclusions}
\label{sec:concl}

In this paper, we have presented a simple scheme for constructing 
models that achieve successful gauge mediation of supersymmetry breaking. 
It uses the essence of the success of the models in our previous 
work~\cite{Murayama:2006yf}, which relies on an approximate $U(1)_R$ 
symmetry for the field that breaks supersymmetry.  We have clarified 
essential ingredients for the scheme: (i) a negative quartic term for 
the supersymmetry breaking field in the K\"ahler potential, (ii) an 
adequate suppression of explicit breaking of the $U(1)_R$ symmetry, 
and (iii) an explicit mass term for the messengers.

We have shown various possible origins for (i).  The negative quartic 
term in the K\"ahler potential may arise at the one-loop level due to 
light fields in the low-energy theory, at tree level in the calculable 
K\"ahler potential along $D$-flat directions, or at the nonperturbative 
level for composite fields.

On general grounds, we need an unexplained suppression of explicit $U(1)_R$ 
breaking terms if the supersymmetry breaking field is an elementary 
singlet.  It could arise accidentally in certain string compactifications 
or due to anthropic reasons on the landscape of theories.  On the other 
hand, models in which a composite field breaks supersymmetry naturally 
suppress the $U(1)_R$ breaking effects and are very attractive.

Finally, the explicit mass term for the messengers may well arise from 
dimensional transmutation at a scale much lower than the cutoff scale, 
due to quantum modified moduli space or gaugino condensations.

We also pointed out that the successful models do rely on small parameters 
to sufficiently suppress gravity-mediated supersymmetry breaking which 
may be flavor non-universal and/or $CP$-violating.  An example is the 
small quark masses in Ref.~\cite{Murayama:2006yf}.  Such small parameters 
again can well arise from dimensional transmutation.

Given a wide variety of classes of models that achieve successful 
gauge mediation, it is also clear that the scheme accommodates a wide 
range of the gravitino mass to alleviate or eliminate cosmological 
problems concerning the gravitino and moduli if any.

We conclude that gauge mediation of supersymmetry breaking is a rather 
generic phenomenon on the landscape of supersymmetric theories.  This 
observation largely eliminates the concern about low-energy supersymmetry 
due to the absence of anomalous flavor-changing and $CP$-violating effects. 
This revitalizes interest in supersymmetry below the TeV scale, which 
will be probed by the forthcoming LHC experiments.

\begin{acknowledgments}
  This work was supported in part by the U.S. DOE under Contract
  DE-AC03-76SF00098, and in part by the NSF under grant PHY-04-57315.
  The work of Y.N. was also supported by the NSF under grant
  PHY-0555661, by a DOE OJI award, and by an Alfred P. Sloan Research
  Fellowship.
\end{acknowledgments}

\end{document}